# QoSIP: A QoS Aware IP Routing Protocol for Multimedia Data

Md. Golam Shagadul Amin Talukder and Al-Mukaddim Khan Pathan*
Department of Computer Science and Engineering, Metropolitan University, Sylhet, Bangladesh
*Department of Computer Science and Information Technology, Islamic University of Technology, Bangladesh.
E-mail: faisal_iut@yahoo.com, mukaddim@iut-dhaka.edu

*Abstract* — Conventional IP routing protocols are not suitable for multimedia applications which have very stringent Quality-of-Service (QoS) demands and they require a connection-oriented service. For multimedia applications it is expected that the router should be able to forward the packet according to the demand of the packet and it is necessary to find a path that satisfies the specific demands of a particular application. In order to address these issues, in this paper, we have presented a QoS aware IP routing protocol where a router stores information about the QoS parameters and routes the packet accordingly.

## 1. Introduction

QoS routing is the routing technique where packet is routed from source to destination selecting the path that satisfies the QoS (i.e., Bandwidth and Delay) requirements. Multimedia data are sensitive to bandwidth and delay. Resource reservation is a necessity for guaranteed end-to-end performance for multimedia applications. However, resource reservation is not well supported in the current routing protocols used in IP network layer. Also, the data packets of these applications could follow different paths and reach the destination out of order, which is not desirable.

Again, multimedia applications demand guaranteed amount of work resources like bandwidth, delay, buffer space, CPU time etc. So, packets need to be routed based on their different QoS requirements, which are not supported by the current IP network. Packets are often routed over paths that are unable to support their requirements, while alternate paths with sufficient resources exist. Hence, there is a need for QoS aware IP routing, which will satisfy the above requirements. The goal of such QoS aware routing algorithm is to find a path in the network that satisfies the given requirements.

In order to address the above-mentioned issues, in this paper, we have proposed a new packet forwarding technique for designing a QoS aware IP routing protocol (QoSIP). In the proposed packet forwarding technique best path will be calculated on the fly whenever a path selection is necessary. For updating state information in the routing table of a router the routers will use "Class based triggering" and the router will maintain state information up to its two-degree neighborhood to bind the flooding. From the performance evaluation of the proposed QoS aware IP routing protocol, it is expected to decrease the computational cost by a large amount and to reduce the protocol overhead to improve the network utilization as well as to fulfill the QoS requirements for packet. The motivation for using a path selection process, based on these requirements is for the hope that it will increase both the services received by users and the overall network efficiency.

The rest of the paper is organized as follows: Section 2 illustrates a review on the related works that have been done on QoS aware IP routing protocols, Section 3 elaborates the proposed QoS aware IP routing protocol (QoSIP), Section 4 presents the simulated result of the proposed QoS aware IP routing protocol using the network simulator NS2, Section 5 performs the performance evaluation of the proposed protocol and Section 6 concludes the paper.

## 2. Related Works

So far, different distributed routing algorithms have been proposed. They can be categorized into two, based on whether all the routers maintain the global state information or not. If the routers maintain the global state information, this information can be used to select a path. But they suffer the common problem of huge overhead to maintain the global state information and in frequently changing environment, this can result state impreciseness. This will not only degrade the performance but also will create routing loops as well. Algorithms proposed by Salama et al. [1] and Su et al. [2] fall into this category.

If no global state is maintained, paths are selected on the fly by using "Flooding". But the overhead of establishing a path is high. Algorithms proposed by Chen et al. [3] and Shin et al. [4] fall into this category.

A third approach used by Ghosh et al. [5] is a combination of the above two techniques. Rather than maintaining global state information, a router stores information only of its immediate neighbors and second-degree neighbors (i.e., neighbors to neighbor). This decreases the overhead and the impreciseness. On the other hand, using information about second-degree neighbors a router forwards packets by bounded two–level flooding.

## 3. Proposed Protocol Architecture

The proposed protocol architecture is divided into two major parts. These are:
- Packet forwarding mechanism
- Routing table maintenance and update policy

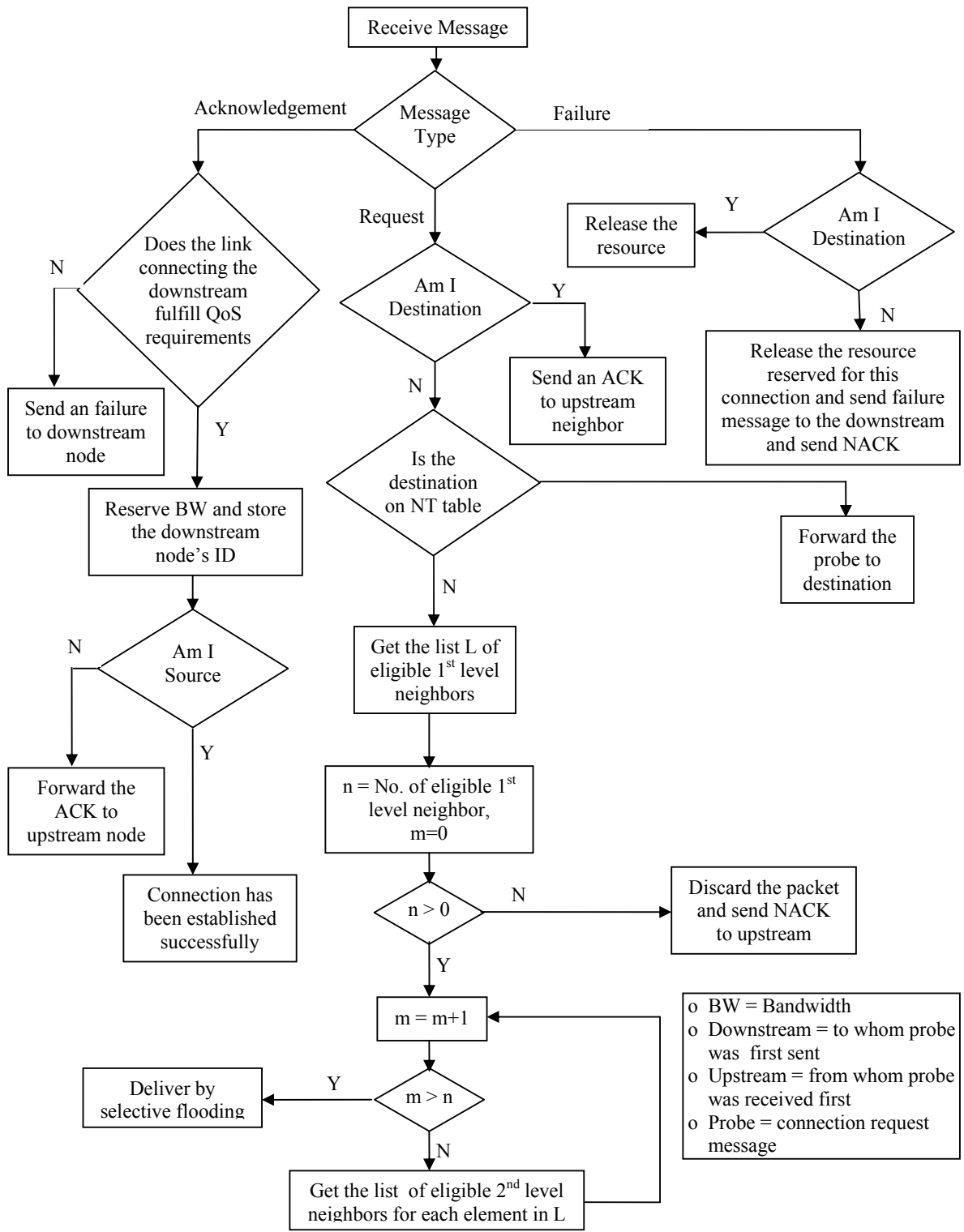

**Figure 1. Flow chart of the routing algorithm**

### 3.1. Packet forwarding mechanism

The packet forwarding mechanism for the proposed routing protocol is described below and a flowchart representing the same mechanism is provided in figure 1.

*3.1.1. Working principle*

Each router maintains a NT (i.e., first degree neighbor table) and a SNT (i.e., a second degree neighbor table). The structures of the tables are described later. A link is said to be

eligible if the link supports QoS requirements. The connection setup process is performed using two phases:
   a. Probing
   b. Acknowledgement
And there is a special case:
   c. Failure handling

The connection setup process along with the connection termination process is described below:

- Connection setup: Whenever a source wants to send any packet, it first sends a probe (i.e., request message for connection) to its neighbor router. Each time a router forwards a packet it will add its address in the packet. Thus, after receiving the packet, destination can know the path and if there are multiple existing paths then it can select the best path. On receiving this probe, each router tries one of the following four alternatives and accordingly handles the packets:
   1. If the router is the destination, then it sends an acknowledgement to the upstream neighbor (i.e., from where the destination received the probe) to establish the connection.
   2. If this router is not the destination, then it checks the NT to find if any of the neighbors is the destination. If it is, then it checks eligibility (i.e., whether it is Bandwidth, Delay supportive) of the link connecting that neighbor. If the link supports requested QoS, then it forwards the packet, otherwise discard the packet and send a negative acknowledgement (NACK) to its upstream.
   3. If the destination router is not in the NT table, then neighbor router checks the SNT table to find whether the destination is one of second-level neighbors of the router. If it is, then it finds the eligible link from SNT corresponding to that destination router and forwards the packet. Otherwise it simply discards the packet and sends a NACK to its upstream.
   4. If the destination is not present in either NT or SNT, it means the destination is beyond the two hops boundary. In this situation, the packet is flooded through all the eligible links via first-degree neighbor towards the second-degree neighbors. It is bounded within the second-degree neighbor.
      4.1. All routers through the paths repeat this forwarding process till the destination gets the probes. The destination may get several feasible paths' (that is, QoS supported path) information. On receiving the packet destination will select the best QoS supported path using Dijkstra [6] or Bellman-Ford [6] shortest path finding algorithm.
      4.2. Then destination will send Acknowledgement (ACK) packet to its best path's upstream router. This is called the ACK phase. The corresponding upstream router will again check the satisfied link that was previously declared by that router. If it satisfies again, then it will reserve resources and store its downstream router's id and forward the packet to its upstream router mentioned by the destination router. Otherwise, it goes to the failure-handling phase by sending a failure message to the downstream router and send NACK to the source using the path mentioned. If this downstream router is not the destination, this will release the resource that it had reserved and send the failure message to its downstream router. This will continue up to the destination.
      4.3. The process mentioned in 4.2 will repeat until source gets the ACK or NACK. In this way connection set up is completed and connection is established.

- Connection termination: The source will send a dummy packet after transmitting the last packet. The dummy packet contains information that instructs each router, which were involved in communication to release reserved resources. Then each router on the way on receiving the dummy packet will release the resources allocated for the link. It will continue up to the destination.

- Failure handling: If any router in the tentative path is unable to reserve the requested bandwidth, it starts the failure-handling phase by sending a failure message to the downstream router. If a router receives a failure message, it releases the resources it had reserved for this connection and forwards the failure message to the next downstream on the tentative path.

3.2. Routing table maintenance

We have assumed that each router keeps:
- The information of all of its neighbors and local metrics in a table called neighbor table (NT).
- The information of all of its second-level neighbors and the metrics to reach them in the second neighbor table (SNT).
- All routers know when to send updates to their neighbors' (i.e. triggering policy).

3.2.1. Building the neighbor table

The neighbor table is a list of all directly connected routers. It gives which link to use to reach a particular neighbor and the resources available along the link. A router can easily build the NT by exchanging HELLO packet with neighbors. Each router on booting up constructs the NT. Table 1 shows the Neighbor Table (NT) of a particular router A in figure 2.

**Table 1. Neighbor Table (NT) of router A**
[D=delay] [B=bandwidth]

| Link  | Neighbor | Available |
|-------|----------|-----------|
| $L_1$ | B        | $D_1B_1$  |
| $L_2$ | C        | $D_2B_2$  |
| …     | …        | …         |

3.2.2. Building second-neighbor table (SNT)

Apart from maintaining an NT, each router also maintains a second-level neighbor table (SNT). This table is a list of routers that are neighbor to the neighbor of a particular router. This SNT also stores the aggregate metrics of the links (two links----one to reach the neighbor and the second one to reach the second neighbor). Figure 2 shows an example network.

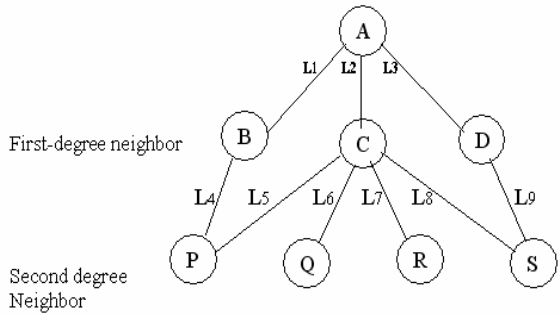

**Figure 2. An Example network**

Let us consider the following for router A in figure 1. In order to construct and maintain the SNT, router must receive updates about the QoS metrics in all its second-level entries. Exchanging special messages called "update" packets does this and the generation of an update packet is determined by the "update policy" described in sub-section 3.3. These "update" packets are constructed by copying the neighbor list and the available QoS metrics from the NT table of the router. All the entries in SNT are added and updated based on "update" packets. Table 2 shows the SNT of router A.

**Table 2. SNT for router A.**

| Link | Neighbor | Second-level neighbor | Available metric |
|---|---|---|---|
| L1, L4 | B | P | D1B1 |
| L2, L6 | C | Q | D2B2 |
| L2, L5 | C | P | D3B3 |
| … | … | … | … |

### 3.3. Update policy

The main idea of sending "update" packet is to communicate the changes in a router's resources availability to other routers. Routers can exchange the "update" packet every time a change occurs in the metric value. But this will certainly increase the network overhead. One solution can be the periodical update, but it is inefficient in QoS routing as the value of the metric changes very frequently. So, within the update interval, the resources can change drastically. If this information is not communicated to other routers, they will have imprecise information and hence the performance of QoS routing will degrade. A detailed survey on various updating policies can be found in [7]. Here we have chosen an updating policy, which is "class based". A particular metric value is divided into adjacent classes and an update is triggered when the current metric value crosses a class boundary. Here metric value will be defined into several class; for example, (0-B), (B-2B), ….etc. After creating NT table the routers will not send any update packet. Rather they will trigger the update when the metrics value crosses a particular class boundary for a link. The router generates the update packets and sends them to all neighbors.

The advantages of using this class based triggering are that router will be able to select an optimum link for a particular application. Say for audio data it needs (0-B) and it is available and (B-2B) is also available. Rather than selecting (B-2B), it will select that link which has class boundary (0-B). Therefore, if further video data come to this router for forwarding, rather than being useless router will be able to reserve that link for video data. So, network utilization is increased. With other triggering methods this achievement is not possible. On the other hand, with the class based triggering, overhead will be decreased since metric value needs to cross a class boundary in order to create a trigger.

### 3.4. Selective flooding

The basic idea of Selective Flooding is to check multiple routes from a source to a destination in parallel before transmitting packet.

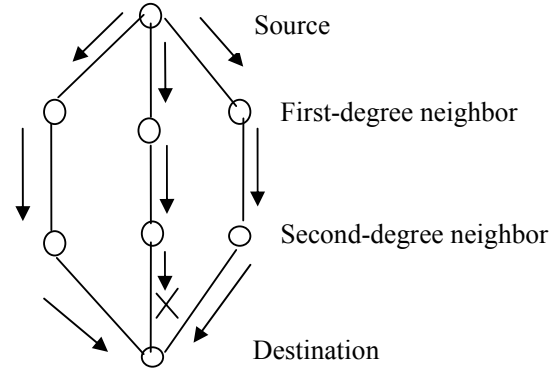

**Figure 3. An example of selective flooding**

In this technique, destination may get several feasible paths. Figure 3 illustrates the Selective Flooding technique. In this figure, destination got two feasible path messages. Among them destination chooses one of the feasible paths and starts communication using that very path. In this case, a router does not contain any link state information. Each router only contains the router list information and corresponding link list for all destinations.

For the proposed routing protocol, we have used the modified version of selecting flooding where the flooding is done only within two-degree neighborhood. If neither first-degree neighbor routers nor second-degree neighbor routers of the source is the destination only then selective flooding is performed.

## 4. Simulation of the Proposed Protocol

The simulation of the proposed protocol is performed using the Network Simulator NS2.

### 4.1. Topology used for simulation

Here the model network topology consists of 15 Routers (depicted on the NAM visualization tool below in figure 4). Each Circle with number inside them indicates a Router. Each Router is connected with full duplex link between them. Each router in the model is QoSIP enabled and CBR has been used as traffic generator. To do experiment, router 0 is considered as source and router 14 as destination for real time communication of the multimedia data. The Red line in the figure 4 shows the most appropriate eligible path (i.e., path that satisfy all the QoS requirements for the packet) that has been selected for the real time communication (according to QoSIP algorithm) among all of the paths before starting communication. All the packets destined to the router 14 follows that path throughout the communication.

4.2. Simulation results

The simulation is run for total 125 minutes. The resultant graph is shown in figure 5

The X axis in the graph represents the time in second and Y axis represents how many bytes received by router 14 (in figure 4) per second. The router 0 starts to send packet towards router 14 when the simulation time is at the point of 27 minute. From the graph it is seen that the router 14 received above 55 Kbytes per second and this amount was almost constant throughout the communication. It is observed that for real time communication of the multimedia data (e.g. video, audio, etc.), if the receiver receives minimum 32 Kbytes per second then the data at the receiver will appear without any flicker. So, according to the obtained result it is possible to construct more than 32 frames [1 frame= 1Kb] at the receiver end whereas 32 frames is the least required criteria for video data.

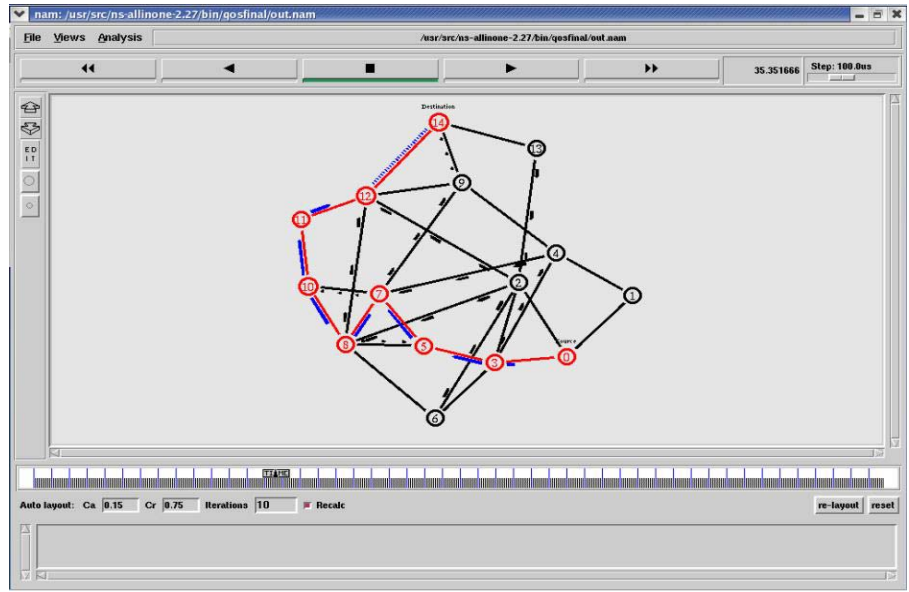

**Figure 4. The topology used for simulation**

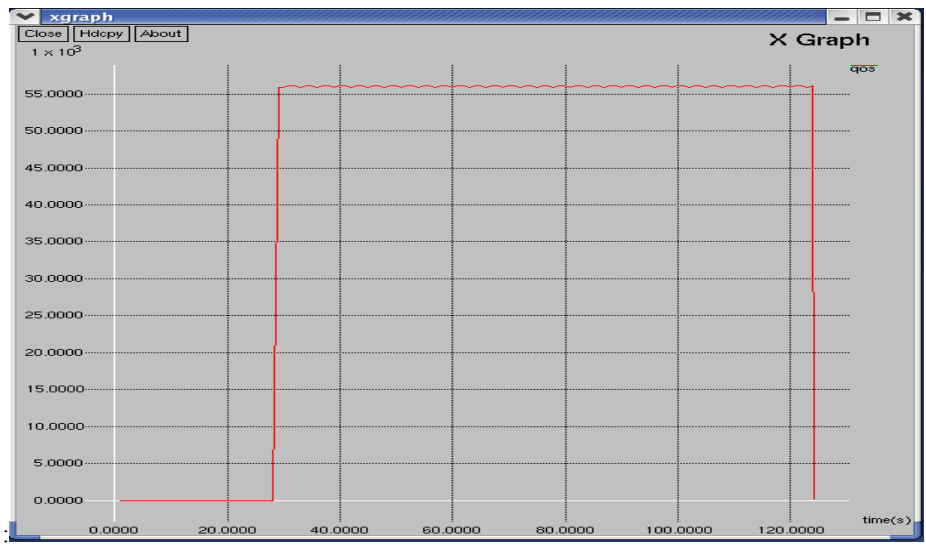

**Figure 5. The resultant graph for the proposed protocol QoSIP**

## 5. Performance Evaluation

The efficiency gain of QoS routing can be measured by comparing the obtained network revenue (e.g., carried traffic) using QoS routing to that of using a simpler routing protocol, which is oblivious to QoS requirements. A QoS aware IP routing protocol route packet based on several parameters and follows a different update policy and path selection criteria. So, a sophisticated computation may take place and it may increase the protocol overhead. We have identified two major components:
- Computational cost
- Protocol overhead

Based on these two components a performance evaluation of the proposed QoS aware IP routing protocol is done as follows:
- Most of the proposed QoS routing protocols require extensive and current knowledge of the entire network in terms of link-state information to compute feasible QoS paths. Selective Flooding eliminates the need for all this information. Only knowledge of the Topology (which is relatively static) is required. Thus the use of selective flooding in the proposed QoS aware IP routing protocol (QoSIP) has enormous potential to reduce the computation at the router and it also reduces the call-blocking rate.
- Again in QoSIP, we have used flooding only within two-degree neighborhood to increase the performance again with the help of SNT or secondary neighbor table. Because in this case flooding is not happened throughout the whole network. So, there will be reduced overhead and traffic flow.
- QoSIP is much better than typical Source Routing. Because source routing stores two important pieces of information at every source. The first one is the network topology (routers and presence of links between them) and the second one is the latest link state information about the whole network. So in this case, transmitting link-state information however is expensive. Such transmission is needed to be done at regular intervals. Thus it consumes non-trivial amounts of bandwidth across the network and uses up considerable computing resources at each router. The more frequent these link-state updates, the more these resources are used. On the other hand, the call-blocking rate increases.
- In QoSIP, we have used NT and SNT that contain limited amount of information only about its first-degree neighborhood and second-degree neighborhood. So, whenever an update is triggered, only limited amount of information is traveled from one router to another. So it reduces the network overhead by reducing the traffic flow. Again maintenance of such table is very easy.
- However, we also need to consider the computation cost at each router when a routing packet arrives. In Source Routing for every link-state update packet each router needs to update its image of the network and recalculate its routing table. This consumes a lot of computation power at the router.
- During Selective Flooding, whenever a control message (probe packet) arrives, all that the router needs to do is to load the necessary QoS information onto the packet and send it to its next hop. Thus even though Selective Flooding sends more routing packets into the network, it saves a lot of precious computation time at the routers.

## 6. Conclusion

In the proposed QoS aware IP routing protocol QoSIP, we have used the selective flooding, which has low overhead when compared to the source routing. On the other hand, the use of class based triggering for updating the table also reduces the overhead, while keeping the metric values updated, which is necessary for forwarding the packet. The proposed IP routing protocol is able find a path that satisfies the QoS requirement if there is any. Furthermore, the proposed protocol can be used for any QoS parameter or for a combination of parameters. It is expected to decrease the computational cost by a large amount and to increase both the services received by users and the overall network efficiency.